\documentstyle{aipproc}

\def\ni{\noindent}  
\def\about{$\sim$} 
\def\Msun{$M_\odot$}
\def\erg/cm2sec{ergs~cm$^{-2}$~s$^{-1}$}   
\def\ergcm2{ergs~cm$^{-2}$}   
\def\cm2{cm$^2$} 
\def\/cm2sec{~cm$^{-2}$~s$^{-1}$}
\def\arcmin{$\,^\prime$~}   
\def\arcsec{$\,^{\prime\prime}$}   
\def\la{\hbox{\rlap{\raise.3ex\hbox{$<$}}\lower.8ex\hbox{$\sim$}\ }}   
\def\ga{\hbox{\rlap{\raise.3ex\hbox{$>$}}\lower.8ex\hbox{$\sim$}\ }}   
\def\deg{$^{\circ}$~}   
   
\def\mdot{$\dot{m}$~}   
\def\X{$\times$~} 
\def\160x40deg{160\deg $\times$ 40\deg}    
\def\20x6deg{20\deg $\times$ 6\deg}    
\def\Ti44{$^{44}$Ti}    
 
\begin{document} 

\title{EXIST: A High Sensitivity Hard X-ray Imaging Sky Survey 
Mission for ISS}
\author{J. Grindlay$^1$, L. Bildsten$^2$, D. Chakrabarty$^3$, 
M. Elvis$^1$, A. Fabian$^4$, F. Fiore$^5$, N. Gehrels$^6$, 
C. Hailey$^7$, F. Harrison$^8$, D. Hartmann$^9$, T. Prince$^8$, 
B. Ramsey$^{10}$, R. Rothschild$^{11}$, G. Skinner$^{12}$, S. Woosley$^{13}$}
\address{$^1$CfA, $^2$ITP/UCSB, $^3$MIT, $^4$IOA/Cambridge, 
$^5$Rome Obs./BeppoSAX, $^6$NASA/GSFC, $^7$Columbia Univ., $^8$Caltech, 
$^9$Clemson Univ., $^{10}$NASA/MSFC, $^{11}$UCSD,\\ 
$^{12}$Birmingham Univ./UK, $^{13}$UC Santa Cruz}

\maketitle

\begin{abstract}
A deep all-sky imaging hard x-ray survey and wide-field 
monitor is needed to extend soft (ROSAT) and medium (ABRIXAS2) 
x-ray surveys into the 10-100 keV band (and beyond) at comparable 
sensitivity (\about0.05 mCrab). This would enable discovery 
and study of \ga3000 obscured AGN, which probably dominate 
the hard x-ray background; detailed study of spectra and 
variability of accreting black holes and a census of BHs 
in the Galaxy; Gamma-ray bursts and associated massive 
star formation (PopIII) at very high redshift and 
Soft Gamma-ray Repeaters throughout the Local Group; and 
a full galactic survey for obscured supernova remnants. 
The Energetic X-ray Imaging Survey Telescope (EXIST) is a  
proposed array of 8 \X 1m$^2$ coded aperture telescopes fixed 
on the International Space Station (ISS) with \160x40deg 
field of view which images the full sky each 90 min orbit. 
EXIST has been included in the most recent NASA Strategic 
Plan as a candidate mission for the next decade.
An overview of the science goals and mission concept is 
presented.

\end{abstract}

\section{Introduction} 
The full sky has not been surveyed in space (imaging) and 
time (variability) at hard x-ray energies. Yet 
the hard x-ray (HX) band, defined here as 10-600 keV, is key to 
some of the most fundamental 
phenomena and objects in astrophysics: the nature and 
ubiquity of active galactic nuclei (AGN), most of 
which are likely to be heavily obscured; the nature and number 
of  black holes; the central engines in 
gamma-ray bursts (GRBs) and the study of GRBs as probes of 
massive star formation in the early 
universe; and the temporal measurement of extremes: from kHz 
QPOs to SGRs for neutron stars, and 
microquasars to Blazars for black holes.

A concept study was conducted for the Energetic X-ray Imaging 
Survey Telescope (EXIST) as one of the 
New Mission Concepts selected in 1994 (Grindlay et al 1995).  
However the rapid pace of discovery in the HX 
domain in the past \about2 years, coupled with 
the promise of a likely 2-10 keV imaging sky survey ABRIXAS2 
(see http://www.aip.de/cgi-bin/w3-msql/groups/xray/abrixas/index.html) 
in c.2002-2004 and the recent selection of Swift 
(see http://swift.gsfc.nasa.gov/) which 
will include a \about10-100 keV partial sky survey 
(to \about1 mCrab) in c.2003-2006, have prompted a much more 
ambitious plan. A dedicated HX survey 
mission is needed  with full sky coverage each orbit and 
\about0.05 mCrab all-sky sensitivity in the 
10-100 keV band (comparable to ABRIXAS2) and extending into the 
100-600 keV band with \about0.5 mCrab 
sensitivity. Such a mission would require very large total 
detector area and large telescope field of view. These needs could 
be met very effectively by a very large 
coded aperture telescope array fixed  (zenith pointing) on the 
International Space Station (ISS), 
and so EXIST-ISS was recommended by the NASA Gamma-Ray Program 
Working Group (GRAPWG) as a 
high priority mission for the coming decade. This mission concept 
has now been included in the NASA 
Strategic Plan formulated in Galveston as a post-2007 candidate mission.
In this paper we summarize the Science Goals and briefly 
present the Mission Concept of EXIST-ISS. Details will 
be presented in forthcoming papers, and are 
partially available on the EXIST website 
(http://hea-www.harvard.edu/EXIST/EXIST.html). 

\section{Science Goals and Objectives} 
EXIST would pursue two key scientific goals: a  
very deep HX imaging and spectral survey,
and a very sensitive HX all-sky variability survey and  
GRB spectroscopy mission. These Survey ({\bf S}) and 
Variability ({\bf V}) goals 
can be achieved by carrying out several primary objectives: \\
 
\ni{\bf S1:~}{\it Sky survey for obscured AGN \& accretion history of universe}\\ 
It is becoming increasingly clear that most of the accretion 
luminosity of the universe is due to obscured 
AGN, and that these objects are very likely the dominant 
sources for the cosmic x-ray (and HX) diffuse 
background (e.g. Fabian 1999). No sky survey has yet been 
carried out to measure the distribution of these 
objects in luminosity, redshift, and broad-band spectra in 
the HX band where, as is becoming increasingly 
clear from BeppoSAX (e.g. Vignati et al 1999),  they are brightest.  
EXIST would detect at least 3000 
Seyfert 2s and conduct a sensitive search for Type 2 QSOs. Spectra 
and variability would be measured, and detailed followup could 
then be carried out with the narrow-field focussing 
HX telescope, HXT, on Constellation-X (Harrison et 
al 1999) as well as IR studies. 
\\

\ni{\bf S2:~}{\it  Black hole accretion on all scales}\\ 
The study of black holes, from x-ray binaries to AGN, 
in the HX band allows their ubiquitous  
Comptonizing coronae to be measured. The relative contributions 
of non-thermal jets at high \mdot 
requires broad band coverage to \ga511 keV, as does the 
transition to ADAFs at lower \mdot values.   
HX spectral variations vs. broad-band flux can test the 
underlying similarities in accretion onto BHs in 
binaries vs. AGN. \\

\ni{\bf S3:~}{\it  Stellar black hole content of Galaxy}\\ 
X-ray novae (XN) appear to be predominantly BH systems, so 
their unbiased detection and sub-arcmin 
locations, which allow optical/IR identifications, can 
provide a direct measure of the 
BH binary content (and XN recurrence time) of the 
Galaxy. XN containing neutron stars can be 
isolated by their usual bursting activity (thermonuclear 
flashes), and since they may solve 
the birth rate problem for millisecond pulsars 
(Yi and Grindlay 1998), their statistics must 
be established. A deep HX survey of the galactic plane
 can also measure the population of 
galactic BHs not in binaries, since they could be 
detected as highly cutoff hard sources 
projected onto giant molecular clouds.  Compared 
to ISM accretion onto isolated 
NSs, for which a few candidates have been found, BHs 
should be much more readily detectable due to 
their intrinsically harder spectra and (much) lower 
expected space velocities, V, and 
larger mass M (Bondi accretion depending on M$^2$/V$^3$). \\ 

\ni{\bf S4:~}{\it  Galactic survey for obscured SNR: SN rate in Galaxy}\\ 
Type II SNe are expected to disperse \about10$^{-4}$ \Msun of \Ti44, with the 
total a sensitive probe of the mass cut and NS formation. 
With a \about87y mean-life 
for decay into $^{44}$Sc which produces narrow lines at 68 and 78 keV, obscured 
SNe can be detected throughout the entire Galaxy for \ga300y given the 
\about10$^{-6}$ photons \/cm2sec line sensitivity and \la2 keV energy 
resolution (at 70 keV) possible for EXIST. Thus 
the likely detection of Cas A (Iyudin et al 1994) can be extended to more distant 
but similarly (or greater) obscured SN to constrain the SN rate in the Galaxy. 
The all-sky imaging of EXIST would extend the central-radian galactic survey 
planned for INTEGRAL to the entire galaxy.\\

\ni{\bf V1:~}{\it  Gamma-ray bursts at the limit: SFR at z \ga10}\\ 
Since at least the ``long'' GRBs located with BeppoSAX are at cosmological 
redshifts, and have apparent luminosities spanning at least a factor of 100, it 
is clear that even the apparently lower luminosity 
GRBs currently detected by BeppoSAX could 
be detected with BATSE out to z \about 4 and 
that the factor \about5 increase in sensitivity with Swift 
will push this back to z \about 5-15 (Lamb and Reichart 1999). 
The additional factor of \about4 increase 
in sensitivity for EXIST would  allow GRB detection and sub-arcmin locations 
for z \ga 15-20 and thus allow the likely epoch of Pop III star formation 
to be probed if indeed GRBs are associated with collapsars (e.g. Woosley 1993) 
produced by the collapse of massive stars. The high throughput and spectral 
resolution for EXIST would enable high time resolution spectra which can 
test internal shock models for GRBs. \\

\ni{\bf V2:~}{\it  Soft Gamma-ray Repeaters: population in Galaxy and Local Group}\\ 
Only 3 SGR sources are known in the Galaxy and 1 in the LMC. Since a
typical \about0.1sec SGR burst spike can be imaged (5$\sigma$) by 
EXIST for a peak flux of \about200 mCrab in the 
10-30 keV band, the typical bursts from the newly discovered
SGR1627-41 (Woods et al 1999) with peak flux \about2 \X 10$^{-6}$ 
erg\/cm2sec would be detected out to \about200 kpc. Hence the brightest 
``normal'' SGR bursts are detectable out to \about3 Mpc and the rare giant 
outbursts (e.g. March 5, 1979 event) out to \about40 Mpc. Thus 
the population and physics of SGRs, and thus their association with 
magnetars and young SNR, can be studied throughout the Local
Group and the rare super-outbursts beyond Virgo. \\

\ni{\bf V3:~}{\it HX blazar alert and spectra: measuring diffuse IR background}\\
The cosmic IR background (CIRB) over \about1-100$\mu$ is poorly measured 
(if at all) and yet can constrain galaxy formation and the luminosity 
evolution of the universe (complementing {\bf S1} above). As reviewed by 
Catanese and Weekes (1999), observing spectral breaks 
(from $\gamma-\gamma$ absorption) for blazars in the 
band \about0.01-100 TeV can measure the CIRB out to z \about1 {\it if} the
intrinsic spectrum is known. Since the $\gamma$-ray spectra of the
detected (low z) blazars are well described by synchrotron-self Compton
(SSC) models, for which the hard x-ray (\about100keV) synchrotron peak
is scattered to the TeV range, the HX spectra can provide both the 
required underlying spectra and time-dependent light curves for all  
objects (variable!) to  be observed with 
GLAST and high-sensitivity ground-based TeV telescopes (e.g. 
VERITAS). \\

\ni{\bf V4:~}{\it  Accretion torques and X-ray pulsars}\\  
The success of BATSE as a HX monitor of bright accreting 
pulsars in the Galaxy (cf. Bildsten et al 1997), in which spin 
histories and accretion torques were derived for a significant 
sample, can be greatly extended with EXIST: the very much larger 
reservoir of Be systems can be explored, and wind vs. disk-fed 
accretion studied in detail. The wide-field HX imaging 
and monitoring capability will also 
allow a new survey for pulsars and AXPs in highly obscured regions 
of the disk, complementing {\bf S4} above. \\

\ni{\bf V5:~}{\it QPOs and accretion disk coronae}\\ 
The rms variability generally and QPO phenomena appear more 
pronounced above \about10keV for x-ray binaries containing both 
BH and NS accretors, suggesting the Comptonizing corona is 
directly involved. Thus QPOs and HX spectral variations can  
allow study of the poorly-understood accretion disk coronae, with 
extension to the AGN case. Although the wide-field increases 
backgrounds, and thus effective modulation, the very large 
area (\ga1m$^2$) of HX imaging area on any given source 
means that  multiple \about100 mCrab 
LMXBs could be simultaneously measured for QPOs with 10\% rms 
amplitude in the poorly explored 10-30 keV band. 

\section{EXIST-ISS Mission Concept}
To achieve the desired \about0.05 mCrab sensitivity full sky up 
to 100 keV (and beyond) requires a very large area array of 
wide-field coded aperture (or other modulation) telescopes. The 
very small field of view (\about10\arcmin) of true focussing  
(e.g. multi-layer) HX telescopes precludes their use for all sky 
imaging and monitoring surveys. EXIST-ISS would take the coded 
aperture concept to a practical limit, with 8 telescopes each 
with 1m$^2$ in effective detector area and 40\deg \X 40\deg 
in field of view (FOV). The individual FOVs are offset by 20\deg 
for a combined FOV of \160x40deg, or \about2sr. By orienting 
the 160\deg axis perpendicular to the orbit vector, the full sky 
can be imaged each orbit if the telescope array is fixed-pointed at 
the local zenith. This gravity-gradient type orientation, and 
the large spatial area of the telescope array, are ideally 
matched for the ISS, which provides a long mounting structure 
(main truss) conveniently oriented perpendicular to the motion, 
as depicted on the EXIST website. 

The sensitivity would yield \about10$^4$ AGNs full sky, thus 
setting a confusion limit resolution requirement 
(\about1/40 ``beam'') of \about5\arcmin. 
With this coded mask pixel size, high energy occulting masks 
(5mm, W) can be constructed with 2.5mm pixel size 
for minimal collimation. The mask shadow is then recorded 
by tiled arrays of CdZnTe (CZT) detectors with effective pixel sizes 
of \about1.3mm, yielding a compact (1.3m) mask-detector spacing. The 
CZT detectors would likely be 20mm square \X 5mm thick (for \ga20\% 
efficiency at 500 keV) and read out by flip-chip bonded 
ASICs (e.g. Bloser et al 1999, Harrison et al 1999). 
 
The 8-telescope array is continuously scanning (sources on 
the orbital plane drift across the 40\deg FOV in 10min; 
correspondingly longer exposures/orbit near the poles), 
with each photon time-tagged and aspect corrected (\about10\arcsec) so that 
ISS pointing errors or flexure are inconsequential over the 
large FOV. Source positions are centroided to \la1\arcmin for 
\ga5$\sigma$ detections. The resulting sky coverage is remarkably 
uniform with \la25\% variation in exposure full sky over the 
\about2mo precession period of the ISS orbit. More details of 
the current mission concept are given in Grindlay et al (2000), 
and will be further developed in the implementation study being 
conducted by the EXIST Science Working Group (EXSWG).


\begin{references}

Bildsten, L. et al, {\it ApJS}, {\bf 113},367 (1997).\\

Bloser, P., Grindlay, J., Narita, T. and Jenkins, J., 
{\it Proc. SPIE}, {\bf 3765}, 388 (1999). \\

Catanese, M. and Weekes, T., {\it PASP}, {\bf 111}, 1193 (1999).\\

Fabian, A. {\it MNRAS}, {\bf 308}, L39 (1999). \\

Grindlay, J.E. et al, {\it Proc. SPIE}, {\bf 2518}, 202 (1995).\\

Grindlay, J.E. et al, {\it Proc. STAIF-2000}, in preparation (2000).\\ 

Harrison, F.A. et al, {\it Proc. SPIE}, {\bf 3765}, 104 (1999). \\

Iyudin, A.F. et al, {\it A\&A}, {\bf 284}, L1 (1994). \\

Lamb, D.Q. and Reichart, D.E., {\it ApJ}, submitted
(astro-ph/9909002) (1999). \\

Vignati, P. et al, {\it A\&A}, {\bf 349}, 57L (1999). \\

Woods, P.M. et al, {\it ApJ}, {\bf 519}, L139 (1999). \\ 

Woosley, S.E. {\it ApJ}, {\bf 405}, 273 (1993). \\

Yi, I. and Grindlay, J.E., {\it ApJ}, {\bf 505}, 828 (1998).\\

\end{references}
\end{document}